\def\etal{{\frenchspacing\it et al.}}

\def\beq#1{\begin{equation}\label{#1}}
\def\eeq{\end{equation}}
\def\beqa#1{\begin{eqnarray}\label{#1}}
\def\eeqa{\end{eqnarray}}

\def\fun#1#2{\lower3.6pt\vbox{\baselineskip0pt\lineskip.9pt
        \ialign{$\mathsurround=0pt#1\hfill##\hfil$\crcr#2\crcr\sim\crcr}}}
        
\def\xi{{{\bf x}^b}}

\documentclass[twocolumn,aps,showpacs,showkeys,nofootinbib]{revtex4}
\usepackage{epsfig}

\newcommand{\be}{\begin{equation}}
\newcommand{\ee}{\end{equation}}
\newcommand{\ba}{\begin{eqnarray}}
\newcommand{\ea}{\end{eqnarray}}

\begin{document}
\input{epsf.sty}

\title{Distance Priors from Planck and Dark Energy Constraints from Current Data}
\author{Yun~Wang$^{1}$\footnote{email: wang@nhn.ou.edu}, 
Shuang Wang$^{1,2}$}
\address{$^{1}$Homer L. Dodge Department of Physics \& Astronomy, Univ. of Oklahoma, 440 W Brooks St., Norman, OK 73019\\
$^{2}$ Department of Physics, College of Sciences, Northeastern University, Shenyang 110004, China}

                 \today

\begin{abstract}

We derive distance priors from Planck first data release, and examine their
impact on dark energy constraints from current observational data.
We give the mean values and covariance matrix of $\{R, l_a, \Omega_b h^2, n_s\}$,
which give an efficient summary of Planck data.
The CMB shift parameters are $R=\sqrt{\Omega_m H_0^2}\,r(z_*)$, 
and $l_a=\pi r(z_*)/r_s(z_*)$, where $z_*$ is the redshift at
the last scattering surface, and $r(z_*)$ and $r_s(z_*)$ denote
our comoving distance to $z_*$ and sound horizon at $z_*$ respectively.

We find that Planck distance priors are significantly tighter
than those from WMAP9.
However, adding Planck distance priors does not lead to significantly
improved dark energy constraints using current data, compared to
adding WMAP9 distance priors. This is because Planck data appear to
favor a higher matter density and lower Hubble constant,
in tension with most of the other current cosmological data sets.
Adding Planck distance priors to current data leads to a
marginal inconsistency with a cosmological constant in a flat universe.

\end{abstract}

%98.80.Es Observational cosmology (including Hubble constant, 
%distance scale, cosmological constant, early universe, etc)
%98.80.-k Cosmology 
%98.80.Jk Mathematical and relativistic aspects of cosmology)


\maketitle

\section{Introduction}

Current observational data do not yet allow us to
differentiate two likely explanations for the observed
cosmic acceleration \cite{Riess98,Perl99}: dark energy, and the modification
of general relativity. For recent reviews, see 
\cite{Copeland06,Ruiz07,Ratra07,Frieman08,Caldwell09,Uzan09,Wang10,Li11,Weinberg12}.
Cosmic acceleration is generally referred to as ``dark energy'' for convenience.

There are three vigorously studied direct probes of dark energy.
Type Ia supernovae (SNe Ia) probe the Hubble parameter $H(z)$ (i.e., the
expansion history of the universe) via the measurement of luminosity distances 
to the SNe Ia \cite{Riess98,Perl99}.
Galaxy clustering (GC) directly probes $H(z)$ (and its integral form $D_A(z)$)
via the baryon acoustic oscillation (BAO) \cite{BG03,SE03} measurements,
and the growth rate $f_g(z)$ (i.e., the growth history of cosmic large scale structure) 
via redshift space distortion measurements.
Weak lensing of galaxies probes a combination of the expansion history
and growth history of the universe \cite{Hu02,Jain03}. 

While these direct probes of cosmic acceleration complement each other,
each with its own set of systematic uncertainties, they require the
inclusion of cosmic microwave background (CMB) anisotropy data to
help break the degeneracies among the dark energy and cosmological parameters.
This is because CMB data provide the strongest constraints on cosmological parameters 
(see, e.g., \cite{Komatsu11}).

Direct measurements of the Hubble constant (see, e.g., \cite{Riess11})
also help break the degeneracy amongst the dark energy and cosmological parameters.
Other data, e.g., gamma ray bursts \cite{Amati02,Bloom03,Schaefer03}, can help 
strengthen the dark energy constraints.

In this paper, we derive distance priors from the Planck first data release,
and examine their impact on dark energy constraints from current
observational data.

We describe our method in Sec.II, present our results in Sec.III,
and conclude in Sec.IV.

\section{Method}
\label{sec:method}

Our main goal is to derive distance priors from Planck data,
and illustrate their impact on current observational data.
For simplicity and clarity,
we only use methods that give geometric constraints 
on dark energy in this paper. The constraints on the growth rate of
cosmic large scale structure are degenerate with the
geometric constraints (see, e.g., \cite{Wang08a,Simpson10}).
We adopt a conservative approach by marginalizing over the growth constraints.

Geometric constraints on dark energy are derived from
the measurement of distances.
The comoving distance to an object at redshift $z$ is given by:
\ba
\label{eq:r(z)}
 & &r(z)=cH_0^{-1}\, |\Omega_k|^{-1/2} {\rm sinn}[|\Omega_k|^{1/2}\, \Gamma(z)]\\
 & &\Gamma(z)=\int_0^z\frac{dz'}{E(z')}, \hskip 1cm E(z)=H(z)/H_0 \nonumber
\ea
where ${\rm sinn}(x)=\sin(x)$, $x$, $\sinh(x)$ for 
$\Omega_k<0$, $\Omega_k=0$, and $\Omega_k>0$ respectively;
and the expansion rate of the universe $H(z)$ (i.e.,
the Hubble parameter) is given by
\ba
\label{eq:H(z)}
&&H^2(z)  \equiv  \left(\frac{\dot{a}}{a}\right)^2 \\
 &= &H_0^2 \left[ \Omega_m (1+z)^3 +\Omega_r (1+z)^4 +\Omega_k (1+z)^2 
+ \Omega_X X(z) \right],\nonumber
\ea
where $\Omega_m+\Omega_r+\Omega_k+\Omega_X=1$, 
$\Omega_m$ includes the contribution from massive neutrinos,
and the dark energy density function $X(z)$ is defined as
\be
X(z) \equiv \frac{\rho_X(z)}{\rho_X(0)}.
\ee
Note that $\Omega_r=\Omega_m /(1+z_{eq}) \ll \Omega_m$ (with $z_{eq}$ denoting
the redshift at matter-radiation equality), thus the $\Omega_r$ term
is usually omitted in dark energy studies at $z\ll 1000$, since dark energy
should only be important at late times.

\subsection{CMB data}
\label{sec:CMB}

CMB data give us the comoving distance to the photon-decoupling surface 
$r(z_*)$, and the comoving sound horizon 
at photon-decoupling epoch $r_s(z_*)$ \cite{Page03}.
Wang \& Mukherjee (2007) \cite{WangPia07} showed that
the CMB shift parameters
\ba
R &\equiv &\sqrt{\Omega_m H_0^2} \,r(z_*)/c, \nonumber\\
l_a &\equiv &\pi r(z_*)/r_s(z_*),
\ea
together with $\omega_b\equiv \Omega_b h^2$, provide an efficient summary
of CMB data as far as dark energy constraints go.
This has been verified by \cite{Li08}.
Replacing $\omega_b$ with $z_*$ gives identical
constraints when the CMB distance priors are
combined with other data \cite{Wang08b}. Using
$\omega_b$, instead of $z_*$, is more appropriate in an MCMC
analysis in which $\omega_b$ is a base parameter.

Note that in order to summarize the CMB data, $R$ and $l_a$ are defined
to contain the physical parameters $\Omega_m h^2$ (matter density), $r(z_*)$ 
(comoving distance to the photon-decoupling surface), and $\theta_* \propto l_a $ 
(angular size of the comoving sound horizon at photon-decoupling epoch),
as these physical parameters are tightly constrained by CMB data and are essentially
independent of model assumptions (except for the assumption on the cosmic curvature).

An intuitive explanation for the effectiveness of $(R, l_a)$ in summarizing
CMB data is as follows.
As indicated by the detailed studies in Wang \& Mukherjee (2007) \cite{WangPia07},
both $R$ and $l_a$ must be used to describe the complex degeneracies amongst the
cosmological parameters that determine the CMB angular power spectrum.
Models that correspond to the same value of $R$ but different
values of $l_a$ give rise to very different CMB angular power
spectra, because $l_a$ determines the average acoustic peak structure.
Models that correspond to the same value of $l_a$ but different values of $R$ have 
the same acoustic peak structure in their CMB angular power spectra, but the
overall amplitude of the acoustic peaks is different in each
model because of the difference in $R$.
The inclusion of CMB lensing data in deriving the shift parameters
makes little difference in their mean values, since the bulk of the 
information comes from the CMB angular power spectra, but
it does reduce their uncertainties by reducing parameter
degeneracies.

The comoving sound horizon at redshift $z$ is given by
\ba
\label{eq:rs}
r_s(z)  &= & \int_0^{t} \frac{c_s\, dt'}{a}
=cH_0^{-1}\int_{z}^{\infty} dz'\,
\frac{c_s}{E(z')}, \nonumber\\
 &= & cH_0^{-1} \int_0^{a} 
\frac{da'}{\sqrt{ 3(1+ \overline{R_b}\,a')\, {a'}^4 E^2(z')}},
\ea
where $a$ is the cosmic scale factor, $a =1/(1+z)$, and
$a^4 E^2(z)=\Omega_m (a+a_{\rm eq})+\Omega_k a^2 +\Omega_X X(z) a^4$,
with $a_{\rm eq}=\Omega_{\rm rad}/\Omega_m=1/(1+z_{\rm eq})$, and
$z_{\rm eq}=2.5\times 10^4 \Omega_m h^2 (T_{CMB}/2.7\,{\rm K})^{-4}$.
The sound speed is $c_s=1/\sqrt{3(1+\overline{R_b}\,a)}$,
with $\overline{R_b}\,a=3\rho_b/(4\rho_\gamma)$,
$\overline{R_b}=31500\Omega_bh^2(T_{CMB}/2.7\,{\rm K})^{-4}$.
We take $T_{CMB}=2.7255$.

The redshift to the photon-decoupling surface, $z_*$, is given by the 
fitting formula \cite{Hu96}:
\be
z_*=1048\, \left[1+ 0.00124 (\Omega_b h^2)^{-0.738}\right]\,
\left[1+g_1 (\Omega_m h^2)^{g_2} \right],
\ee
where
\ba
g_1 &= &\frac{0.0783\, (\Omega_b h^2)^{-0.238}}
{1+39.5\, (\Omega_b h^2)^{0.763}}\\
g_2 &= &\frac{0.560}{1+21.1\, (\Omega_b h^2)^{1.81}}
\ea
The redshift of the drag epoch $z_d$ is well approximated by \cite{EisenHu98}
\begin{equation}
z_d  =
 \frac{1291(\Omega_mh^2)^{0.251}}{1+0.659(\Omega_mh^2)^{0.828}}
\left[1+b_1(\Omega_bh^2)^{b2}\right],
\label{eq:zd}
\end{equation}
where
\begin{eqnarray}
  b_1 &= &0.313(\Omega_mh^2)^{-0.419}\left[1+0.607(\Omega_mh^2)^{0.674}\right],\\
  b_2 &= &0.238(\Omega_mh^2)^{0.223}.
\end{eqnarray}

\begin{figure} 
\psfig{file=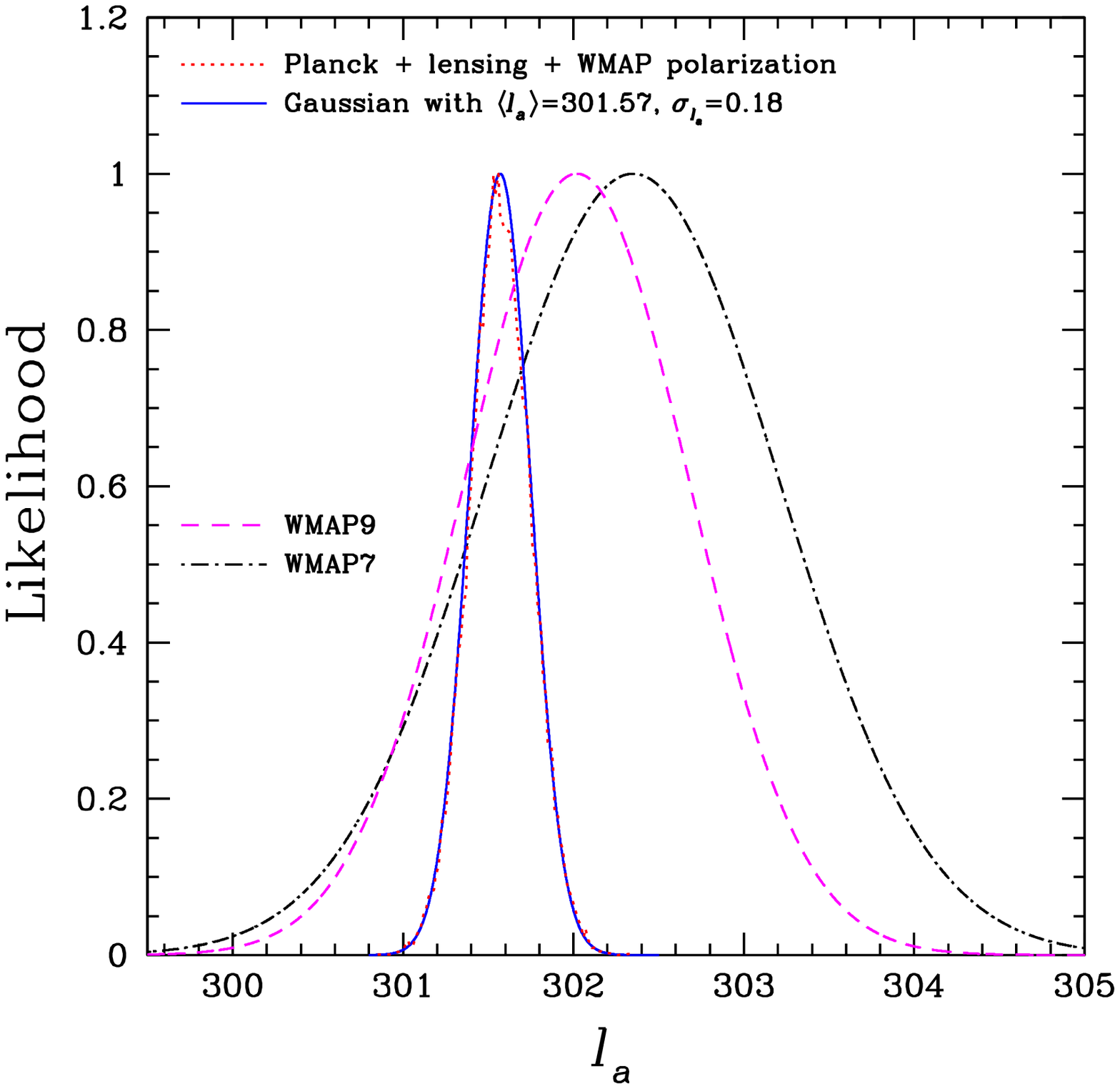,width=3in}\\
\vspace{-0.4in}
\caption{\label{fig:l_a}\footnotesize%
One-dimensional marginalized probability distributions
of CMB shift parameter $l_a$ derived from Planck, WMAP9, and WMAP7 data.}
\end{figure}

\begin{figure} 
\psfig{file=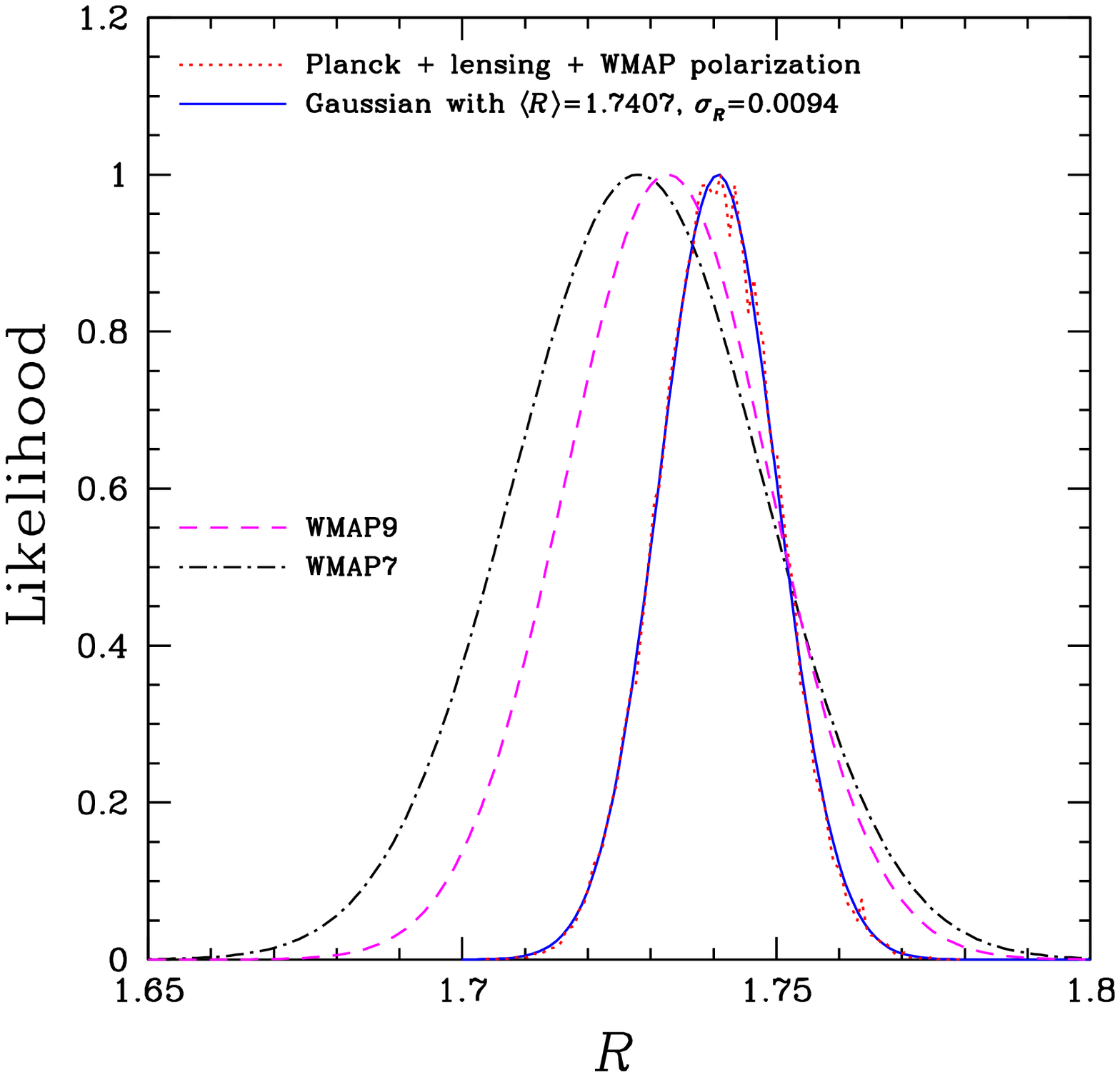,width=3in}\\
\vspace{-0.4in}
\caption{\label{fig:R}\footnotesize%
One-dimensional marginalized probability distribution 
of CMB shift parameter $R$ derived from Planck, WMAP9, and WMAP7 data.
}
\end{figure}

\begin{figure} 
\psfig{file=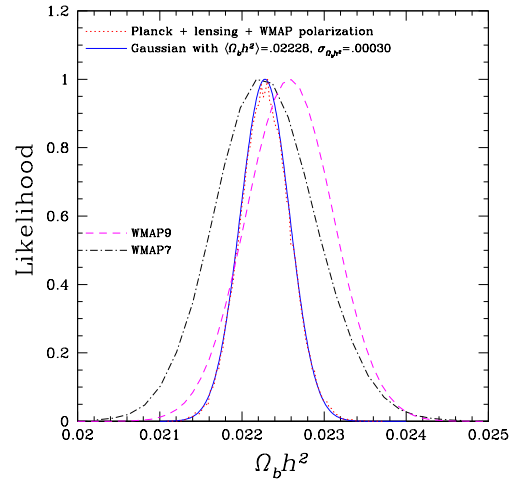,width=3in}\\
\vspace{-0.4in}
\caption{\label{fig:obh2}\footnotesize%
One-dimensional marginalized probability distribution 
of the dimensionless baryon density, $\omega_b\equiv \Omega_b h^2$, 
derived from Planck, WMAP9, and WMAP7 data.
}
\end{figure}

\begin{figure} 
\psfig{file=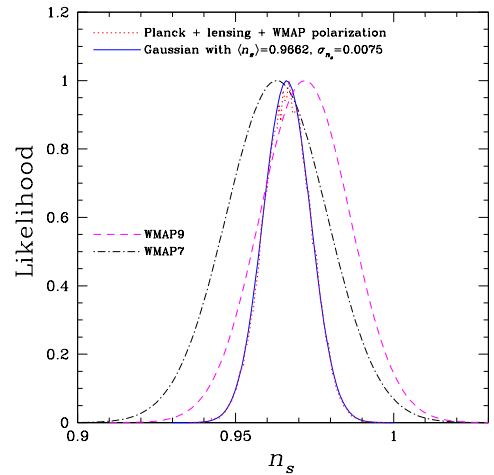,width=3in}\\
\vspace{-0.4in}
\caption{\label{fig:ns}\footnotesize%
One-dimensional marginalized probability distributions
of powerlaw index of primordial matter power spectrum, $n_s$,
derived from Planck, WMAP9, and WMAP7 data.
}
\end{figure}

Figs.\ref{fig:l_a}-\ref{fig:ns} show the one-dimensional marginalized probability 
distributions (pdf) of $(l_a, R, \omega_b, n_s)$ from Planck \cite{planckXVI}, 
WMAP9 \cite{Bennett12}, and WMAP7 \cite{Komatsu11} data,
for $w_X(z)=-1$ and one massive neutrino (with mass of 0.06eV), 
and without assuming a flat universe. We have used the Planck archiv
data to obtain constraints on $(l_a, R, \omega_b, n_s)$ from Planck and WMAP9; 
this archiv does not include $w_X(z)\neq -1$ cases without assuming a flat universe.
Fortunately, the constraints on $(l_a, R, \omega_b, n_s)$ 
(including the pdf's) are {\it not} sensitive to the assumption about dark energy \cite{Wang12CM}.
Three sets of pdf's are shown in Figs.\ref{fig:l_a}-\ref{fig:ns}:\\
\noindent
(1) Planck+lensing+WP: Planck temperature data combined with Planck lensing, as well as
WMAP polarization at low multipoles ($l \leq 23$). This set represents the tightest constraints
from CMB data only at present.
Note that excluding the Planck lensing data changes the mean values of $(l_a, R, \omega_b, n_s)$
by 0.2\% or less, and increases the dispersions slightly.\\
\noindent
(2) WMAP: WMAP 9 year temperature and polarization data.\\
\noindent
(3) WMAP7: WMAP 7 year temperature and polarization data.

The Planck+lensing+WP pdf's in Figs.\ref{fig:l_a}-\ref{fig:ns} are well fitted by Gaussian
distributions with the following means and standard deviations:
\ba
&&\langle l_a \rangle = 301.57, \sigma(l_a)=0.18 \nonumber\\
&&\langle R \rangle = 1.7407,  \sigma(R)=0.0094\nonumber\\
&& \langle \omega_b \rangle = 0.02228, \sigma(\omega_b)=0.00030\nonumber\\
&& \langle n_s \rangle = 0.9662, \sigma(n_s)=0.0075.
\label{eq:CMB_mean_planck}
\ea
The normalized covariance matrix of $(l_a, R, \omega_b, n_s)$ is
\be
\left(
\begin{array}{cccc} 
   1.0000  &    0.5250  &   -0.4235  &   -0.4475    \\
  0.5250  &     1.0000  &   -0.6925  &   -0.8240    \\
 -0.4235  &   -0.6925  &     1.0000  &    0.6109    \\
 -0.4475 &   -0.8240  &    0.6109 &     1.0000  \\
\end{array}
\right)
\label{eq:normcov_planck}
\ee

For comparison, we have also obtained the WMAP9 constraints
on $(l_a, R, \omega_b, n_s)$. The WMAP9 in Figs.\ref{fig:l_a}-\ref{fig:ns} are well fitted by Gaussian
distributions with the following means and standard deviations:
\ba
&&\langle l_a \rangle = 302.02, \sigma(l_a)=0.66 \nonumber\\
&&\langle R \rangle = 1.7327,  \sigma(R)=0.0164\nonumber\\
&& \langle \omega_b \rangle = 0.02260, \sigma(\omega_b)=0.00053\nonumber\\
&& \langle n_s \rangle = 0.9719, \sigma(n_s)=0.0143.
\label{eq:CMB_mean_wmap9}
\ea
The normalized covariance matrix of $(l_a, R, \omega_b, n_s)$ is
\be
\left(
\begin{array}{cccc} 
   1.0000  &    0.3883   &  -0.6089  &   -0.5391   \\
  0.3883   &    1.0000   &  -0.5239 &   -0.6523    \\
 -0.6089   &  -0.5239   &    1.0000  &    0.8563    \\
 -0.5391   &  -0.6523  &   0.8563 &     1.0000   \\
\end{array}
\right)
\label{eq:normcov_wmap9}
\ee
The WMAP7 constraints are from \cite{Wang12CM}.

Since the primary GC data we use in this paper have been marginalized 
over $n_s$ \cite{CW12}, we should marginalize the CMB distance priors 
over $n_s$ as well.\footnote{An improved approach is to use GC data
that retain the $n_s$ dependence, and combine with the CMB constraints
on $(l_a, R, \omega_b, n_s)$.}
This means dropping the 4th row and 4th column from
the normalized covariance matrix of $(l_a, R, \omega_b, n_s)$,
then obtain the covariance matrix for $(l_a, R, \omega_b)$ as follows:
\be
\mbox{Cov}_{CMB}(p_i,p_j)=\sigma(p_i)\, \sigma(p_j) \,\mbox{NormCov}_{CMB}(p_i,p_j),
\label{eq:CMB_cov}
\ee
where $i,j=1,2,3$. The rms variance $\sigma(p_i)$ and the normalized
covariance matrix $\mbox{NormCov}_{CMB}$ are given by
Eqs.(\ref{eq:CMB_mean_planck}) and (\ref{eq:normcov_planck}) for
Planck+lensing+WP, and Eqs.(\ref{eq:CMB_mean_wmap9}) and (\ref{eq:normcov_wmap9})
for WMAP9 respectively.

CMB data are included in our analysis by adding
the following term to the $\chi^2$ of a given model
with  $p_1=l_a(z_*)$, $p_2=R(z_*)$,and $p_3=\omega_b$:
\be
\label{eq:chi2CMB}
\chi^2_{CMB}=\Delta p_i \left[ \mbox{Cov}^{-1}_{CMB}(p_i,p_j)\right]
\Delta p_j,
\hskip .5cm
\Delta p_i= p_i - p_i^{data},
\ee
where $p_i^{data}$ are the mean from Eq.(\ref{eq:CMB_mean_planck}) and
Eq.(\ref{eq:CMB_mean_wmap9}),
and ${\rm Cov}^{-1}_{CMB}$ is the inverse of the covariance matrix of 
[$l_a(z_*), R(z_*),  \omega_b]$ from Eq.(\ref{eq:CMB_cov}).
Note that $p_4=n_s$ should be added if the constraints on
$n_s$ are included in the GC data.

\subsection{Analysis of SN Ia Data}

SN Ia data give measurements of the luminosity distance 
$d_L(z)$ through that of the distance modulus
of each SN:
\be
\label{eq:m-M}
\mu_0 \equiv m-M= 5 \log\left[\frac{d_L(z)}{\mathrm{Mpc}}\right]+25,
\ee
where $m$ and $M$ represent the apparent and absolute magnitude
of a SN. The luminosity distance $d_L(z)=(1+z)\, r(z)$, with the comoving
distance $r(z)$ given by Eq.(\ref{eq:r(z)}). 

Care must be taken in interpreting supernova distances in an inhomogeneous
universe \cite{Clarkson11}.
We use the compilation of SN Ia data by Conley et al. (2011) \cite{Conley11}, 
which include the SNe Ia from the first three years of the Supernova Legacy 
Survey (SNLS3), the largest homogeneous SN Ia data set publicly available at present, and apply
flux-averaging to reduce the systematic bias due to weak lensing magnification
of SNe \cite{Wang00,WangPia04,Wang05}, as detailed in Wang, Chuang, \& Mukherjee (2012) \cite{Wang12CM}.

For a set of 472 SNe Ia, Conley et al. (2011) \cite{Conley11} give 
the apparent $B$ magnitude, $m_B$, and the covariance matrix for
$\Delta m \equiv m_B-m_{\rm mod}$, with
\be
m_{\rm mod}=5 \log_{10}{\cal D}_L(z|\mbox{\bf s})
- \alpha (s-1) +\beta {\cal C} + {\cal M},
\ee
where ${\cal D}_L(z|\mbox{\bf s})$ is the luminosity distance
multiplied by $H_0$
for a given set of cosmological parameters $\{ {\bf s} \}$,
$s$ is the stretch measure of the SN light curve shape, and
${\cal C}$ is the color measure for the SN.
${\cal M}$ is a nuisance parameter representing some combination
of the absolute magnitude of a fiducial SN Ia, $M$, and the 
Hubble constant $H_0$.
%mu0pa(j)=5.*log10((1.+zhel(j))*rcov0)-alpha*(str(j)-1.0)+beta*color(j)+Msn
Since the time dilation part of the observed luminosity distance depends 
on the total redshift $z_{\rm hel}$ (special relativistic plus cosmological),
we have \cite{Hui06}
\be
{\cal D}_L(z|\mbox{\bf s})\equiv c^{-1}H_0 (1+z_{\rm hel}) r(z|\mbox{\bf s}),
\ee
where $z$ and $z_{\rm hel}$ are the CMB restframe and heliocentric redshifts
of the SN. 

For a set of $N$ SNe with correlated errors, we have \cite{Conley11}
\be
\label{eq:chi2_SN}
\chi^2=\Delta \mbox{\bf m}^T \cdot \mbox{\bf C}^{-1} \cdot \Delta\mbox{\bf m}
\ee
where $\Delta \bf m$ is a vector with $N$ components, and
$\mbox{\bf C}$ is the $N\times N$ covariance matrix of the SNe Ia.

Note that $\Delta m$ is equivalent to $\Delta \mu_0$, since 
\be
\Delta m \equiv m_B-m_{\rm mod}
= \left[m_B+\alpha (s-1) -\beta {\cal C}\right] - {\cal M}.
\ee

The total covariance matrix is \cite{Conley11}
\be
\mbox{\bf C}=\mbox{\bf D}_{\rm stat}+\mbox{\bf C}_{\rm stat}
+\mbox{\bf C}_{\rm sys},
\ee
with the diagonal part of the statistical uncertainty given by \cite{Conley11}
\ba
\mbox{\bf D}_{\rm stat,ii}&=&\sigma^2_{m_B,i}+\sigma^2_{\rm int}
+ \sigma^2_{\rm lensing}+ \sigma^2_{{\rm host}\,{\rm correction}} \nonumber\\
&& + \left[\frac{5(1+z_i)}{z_i(1+z_i/2)\ln 10}\right]^2 \sigma^2_{z,i} 
 +\alpha^2 \sigma^2_{s,i}+\beta^2 \sigma^2_{{\cal C},i} \nonumber\\
&& + 2 \alpha C_{m_B s,i} - 2 \beta C_{m_B {\cal C},i}
-2\alpha\beta C_{s {\cal C},i},
\ea
where $C_{m_B s,i}$, $C_{m_B {\cal C},i}$, and $C_{s {\cal C},i}$
are the covariances between $m_B$, $s$, and ${\cal C}$ for the $i$-th SN. 
Note also that $\sigma^2_{z,i}$ includes a
peculiar velocity residual of 0.0005 (i.e., 150$\,$km/s) added 
in quadrature \cite{Conley11}.

The statistical and systematic covariance matrices, 
$\mbox{\bf C}_{\rm stat}$ and $\mbox{\bf C}_{\rm sys}$,
are generally not diagonal \cite{Conley11}, and are given in the
form:
\be
\mbox{\bf C}_{\rm stat}+\mbox{\bf C}_{\rm sys}
=V_0+\alpha^2 V_a + \beta^2 V_b + 2 \alpha V_{0a}
-2 \beta V_{0b} - 2 \alpha\beta V_{ab}.
\ee
where $V_0$, $V_{a}$, $V_{b}$, $V_{0a}$, $V_{0b}$, and
$V_{ab}$ are matrices given by the SNLS data archive
at https://tspace.library.utoronto.ca/handle/1807/24512/.
$\mbox{\bf C}_{\rm stat}$ includes the uncertainty in
the SN model. $\mbox{\bf C}_{\rm sys}$ includes the
uncertainty in the zero point. Note that  $\mbox{\bf C}_{\rm stat}$
and $\mbox{\bf C}_{\rm sys}$ do not depend on ${\cal M}$, 
since the relative distance moduli are independent of the value 
of ${\cal M}$ \cite{Conley11}.

We refer the reader to Conley et al. (2011) \cite{Conley11}
for a detailed discussion of the origins of the statistical
and systematic errors. As an example, we note that the correlation 
of errors on different SNe arises from a statistical uncertainty 
in the zero point of one passband, e.g., $r_M$.
This directly affects all SNe with $r_M$ measurements
due to K-corrections (restframe $B$ to $r_M$), and
indirectly affects even the SNe without $r_M$ measurements 
through the empirical SN models by changing
the templates and the measured color-luminosity relationship.

For $\chi^2$ statistics using MCMC or a grid of parameters, 
here are the steps in flux-averaging \cite{Wang12CM}:

(1) Convert the distance modulus of SNe Ia into 
``fluxes'',
\be
\label{eq:flux}
F(z_l) \equiv 10^{-(\mu_0^{\rm data}(z_l)-25)/2.5} =  
\left( \frac{d_L^{\rm data}(z_l)} {\mbox{Mpc}} \right)^{-2}.
\ee

(2) For a given set of cosmological parameters $\{ {\bf s} \}$,
obtain ``absolute luminosities'', \{${\cal L}(z_l)$\}, by
removing the redshift dependence of the ``fluxes'', i.e.,
\be
\label{eq:lum}
{\cal L}(z_l) \equiv d_L^2(z_l |{\bf s})\,F(z_l).
\ee

(3) Flux-average the ``absolute luminosities'' \{${\cal L}^i_l$\} 
in each redshift bin $i$ to obtain $\left\{\overline{\cal L}^i\right\}$:
\be 
 \overline{\cal L}^i = \frac{1}{N_i}
 \sum_{l=1}^{N_i} {\cal L}^i_l(z^{(i)}_l),
 \hskip 1cm
 \overline{z_i} = \frac{1}{N_i}
 \sum_{l=1}^{N_i} z^{(i)}_l. 
\ee

(4) Place $\overline{\cal L}^i$ at the mean redshift $\overline{z}_i$ of
the $i$-th redshift bin, now the binned flux is
\be
\overline{F}(\overline{z}_i) = \overline{\cal L}^i /
d_L^2(\overline{z}_i|\mbox{\bf s}).
\ee

(5) Compute the covariance matrix of $\overline{F}(\overline{z}_i)$
and $\overline{F}(\overline{z}_j)$:
\ba
&& \mbox{Cov}\left[\overline{F}(\overline{z}_i),\overline{F}(\overline{z}_j)\right] \\
&=&\frac{1}{N_i N_j}\left[\frac{\ln 10 /2.5}
{d_L(\overline{z}_i|\mbox{\bf s})d_L(\overline{z}_j|\mbox{\bf s})}
\right]^2 \cdot \nonumber\\
&& \sum_{l=1}^{N_i} \sum_{m=1}^{N_j} {\cal L}(z_l^{(i)})
{\cal L}(z_m^{(j)}) \langle \Delta \mu_0^{\rm data}(z_l^{(i)})\Delta 
\mu_0^{\rm data}(z_m^{(j)})
\rangle \nonumber 
\ea
where $\langle \Delta \mu_0^{\rm data}(z_l^{(i)})\Delta \mu_0^{\rm data}(z_m^{(j)})\rangle $
is the covariance of the measured distance moduli of the $l$-th SN Ia
in the $i$-th redshift bin, and the $m$-th SN Ia in the $j$-th
redshift bin. ${\cal L}(z)$ is defined by Eqs.(\ref{eq:flux}) and (\ref{eq:lum}).

(6) For the flux-averaged data, $\left\{\overline{F}(\overline{z}_i)\right\}$, 
compute
\be
\label{eq:chi2_SN_fluxavg}
\chi^2 = \sum_{ij} \Delta\overline{F}(\overline{z}_i) \,
\mbox{Cov}^{-1}\left[\overline{F}(\overline{z}_i),\overline{F}(\overline{z}_j)
\right] \,\Delta\overline{F}(\overline{z}_j)
\ee
where
\be
\Delta\overline{F}(\overline{z}_i) \equiv 
\overline{F}(\overline{z}_i) - F^p(\overline{z}_i|\mbox{\bf s}),
\ee
with $F^p(\overline{z}_i|\mbox{\bf s})=
\left( d_L(z|\mbox{\bf s}) /\mbox{Mpc} \right)^{-2}$.

For the sample of SNe we use in this study, we  
flux-averaged the SNe with $dz=0.04$, to ensure that
almost all redshift bins contain at least one SN.
Our SN flux-averaging code is available 
at http://www.nhn.ou.edu/$\sim$wang/SNcode/.

\subsection{Galaxy Clustering Data}
\label{sec:GC}

For GC data, we use the measurements of $H(z)r_s(z_d)/c$ and $D_A(z)/r_s(z_d)$ 
(where $H(z)$ is the Hubble parameter, $D_A(z)$ is the angular diameter distance, 
and $r_s(z_d)$ is the sound horizon at the drag epoch) from the two-dimensional two-point correlation
function measured at z=0.35 \cite{CW12} and z=0.57 \cite{C13}.
The $z=0.35$ measurement was made by Chuang \& Wang (2012) \cite{CW12} using
a sample of the SDSS DR7 Luminous Red Galaxies (LRGs).
The $z=0.57$ measurement was made by Chuang et al. (2013) \cite{C13} using the
CMASS galaxy sample from BOSS.  

Using the two-dimensional two-point correlation function of SDSS DR7
in the scale range of 40-120$\,$Mpc/$h$, 
Chuang \& Wang (2012) \cite{CW12} found that
\ba
H(z=0.35)r_s(z_d)/c&=&0.0434  \pm  0.0018  \nonumber \\
D_A(z=0.35)/r_s(z_d)&=& 6.60  \pm  0.26\nonumber\\
r&=&0.0604
\label{eq:CW2}
\ea
where $r$ is the normalized correlation coefficient between
$H(z=0.35)r_s(z_d)/c$ and $D_A(z=0.35)/r_s(z_d)$, and 
$r_s(z_d)$ is the sound horizon at the drag epoch (given by
Eqs.(\ref{eq:rs}) and (\ref{eq:zd}). 

In a similar analysis using the CMASS galaxy sample from BOSS, 
Chuang et al. (2013) found that
\ba
H(z=0.57)r_s(z_d)/c&=&0.0454	\pm  0.0031 \nonumber \\
D_A(z=0.57)/r_s(z_d)&=& 8.95	\pm  0.27 \nonumber\\
r&=&0.4874
\label{eq:C13}
\ea
We marginalize over the growth rate measurement made by Chuang et al. 2013
\cite{C13} for a conservative approach.

GC data are included in our analysis by adding $\chi^2_{GC}=\chi^2_{GC1}+\chi^2_{GC2}$,
with $z_{GC1}=0.35$ and $z_{GC2}=0.57$, to the $\chi^2$ of a given model.
Note that
\be
\label{eq:chi2bao}
\chi^2_{GCi}=\Delta p_i \left[ {\rm C}^{-1}_{GC}(p_i,p_j)\right]
\Delta p_j,
\hskip .5cm
\Delta p_i= p_i - p_i^{data},
\ee
where $p_1=H(z_{GCi})r_s(z_d)/c$ and $p_2=D_A(z_{GCi})/r_s(z_d)$,
with $i=1,2$.
 
\subsection{Gammay-ray Burst Data}
\label{sec:GRB}

We add gammay-ray burst (GRB) data to our analysis, since these are
complementary in redshift range to the SN Ia data.
We use GRB data in the form of the model-independent GRB distance 
measurements from Wang (2008c) \cite{Wang08c}, which were
derived from the data of 69 GRBs with $0.17 \le z \le 6.6$
from Schaefer (2007) \cite{Schaefer07}\footnote{The proper
calibration of GRBs is an active area of research. For 
recent studies on the impact of detector thresholds, 
spectral analysis, and unknown selection effects, see, e.g., 
\cite{Butler07,Butler09,Petrosian09,Shahmoradi11}.}.

The GRB distance measurements are given in terms of \cite{Wang08c}
\be
\label{eq:rp}
\overline{r_p}(z_i)\equiv \frac{r_p(z)}{r_p(0.17)}, \hskip 1cm
r_p(z) \equiv \frac{(1+z)^{1/2}}{z}\, \frac{H_0}{ch}\, r(z),
\ee
where $r(z)$ is the comoving distance at $z$.

The GRB data are included in our analysis by adding
the following term to the $\chi^2$ of a given model:
\ba
\label{eq:rGRB1}
\chi^2_{GRB} &= & \left[\Delta \overline{r_p}(z_i)\right]  \cdot
\left(\mathrm{Cov}^{-1}_{GRB}\right)_{ij}\cdot
\left[\Delta \overline{r_p}(z_j)\right]
\nonumber\\
\Delta \overline{r_p}(z_i) &= & \overline{r_p}^{\mathrm{data}}(z_i)-\overline{r_p}(z_i),
\ea
where $\overline{r_p}(z)$ is defined by Eq.(\ref{eq:rp}).
The covariance matrix is given by
\be
\left(\mathrm{Cov}_{GRB}\right)_{ij}=
\sigma(\overline{r_p}(z_i)) \sigma(\overline{r_p}(z_j)) 
\left(\overline{\mathrm{Cov}}_{GRB}\right)_{ij},
\ee
where $\overline{\mathrm{Cov}}_{GRB}$ is the normalized covariance matrix
from Table 3 of Wang (2008c) \cite{Wang08c}, and
\ba
\label{eq:rGRB3}
\sigma(\overline{r_p}(z_i))  &= &\sigma\left(\overline{r_p}(z_i)\right)^+, \hskip 0.5cm \mathrm{if}\,\, 
\overline{r_p}(z) \ge \overline{r_p}(z)^{\mathrm{data}}; \nonumber\\
\sigma(\overline{r_p}(z_i))  &= &\sigma\left(\overline{r_p}(z_i)\right)^-, \hskip 0.5cm \mathrm{if}\,\, 
\overline{r_p}(z) < \overline{r_p}(z)^{\mathrm{data}},
\ea
where $\sigma\left(\overline{r_p}(z_i)\right)^+$ and 
$\sigma\left(\overline{r_p}(z_i)\right)^-$ are the 68\% C.L. errors
given in Table 2 of Wang (2008c) \cite{Wang08c}.

\subsection{Dark energy parametrization}
\label{sec:para}

Since we are ignorant of the true nature of dark energy,
it is useful to measure the dark energy density function
$X(z)\equiv \rho_X(z)/\rho_X(0)$ as a free function of
redshift \cite{WangGarnavich,WangTegmark04,WangFreese06}.
This has the advantage of allowing dark energy
models in which $\rho_X(z)$ becomes negative in the future,
e.g., the ``Big Crunch'' models \cite{Linde87,WangLinde04},
which are precluded if we parametrize dark energy with 
an equation of state $w_X(z)$ \cite{WangTegmark04}.

Here we parametrize $X(z)$ by cubic-splining its values
at $z=1/3$, $2/3$, and 1.0, and assume that
$X(z>1)=X(z=1)$. For simplicity of notation,
we define $X_{0.33}\equiv X(z=1/3)$,
$X_{0.67}\equiv X(z=2/3)$, and $X_{1.0}\equiv X(z=1)$.
Fixing $X(z>1)$ reflects the limit
of current data, and avoids making assumptions about early
dark energy that can be propagated into artificial constraints on
dark energy at low $z$ \cite{WangTegmark04,WangPia07}.

For comparison with the work of others, we also 
consider a dark energy equation of state linear in the cosmic scale 
factor $a$, $w_X(a)=w_0+(1-a)w_a$ \cite{Chev01}.
A related parametrization is \cite{Wang08b}
\be
w_X(z)=w_0(3a-2)+ 3w_{0.5}(1-a),
\ee
where $w_{0.5}\equiv w_X(z=0.5)$.
Wang (2008b) \cite{Wang08b} showed that ($w_0$, $w_{0.5}$) are much less
correlated than ($w_0$, $w_a$), thus are a better set of parameters
to use. We find that ($w_0$, $w_{0.5}$) converge much faster
than ($w_0$, $w_a$) in a Markov Chain Monte Carlo (MCMC)
likelihood analysis for the same data.

\section{Results}

We perform a MCMC likelihood analysis \cite{Lewis02} to obtain 
${\cal O}$($10^6$) samples for each set of results presented in 
this paper. We assume flat priors for all the parameters, and allow ranges 
of the parameters wide enough such that further increasing the allowed 
ranges has no impact on the results. We process the MCMC chains
following the standard practice of ensuring convergence and
thinning using CosmoMC.

In addition to the SN Ia, CMB, GC, and GRB data discussed in 
Sec.{\ref{sec:method}}, we impose a prior of 
$H_0 = 73.8 \pm 2.4\,$km$\,$s$^{-1}$Mpc$^{-1}$, from the
HST measurements by Riess et al. (2011) \cite{Riess11}.

We do {\it not} assume a flat universe.
In addition to the dark energy parameters described in Sec.\ref{sec:para},
we also constrain cosmological parameters ($\Omega_m, \Omega_k, h, \omega_b$), 
where $\omega_b\equiv \Omega_b h^2$. In addition, we marginalize over
the SN Ia nuisance parameters $\{\alpha, \beta, {\cal M}\}$.
We only use flux-averaged SN Ia data (with ${\rm d}z=0.04$), as flux-averaging reduces
the impact of systematic uncertainties on dark energy and cosmological
parameter constraints \cite{Wang12CM}.

We will present results for dark energy density at
$z=1/3$, $2/3$, and 1, as well as ($w_0,w_a$) and 
($w_0,w_{0.5}$),
and a constant dark energy equation of state $w$.

%\subsection{Comparison of Planck and WMAP9 Distance Priors}
\subsection{Constraints on dark energy density function $X(z)$}

\begin{figure} 
\psfig{file=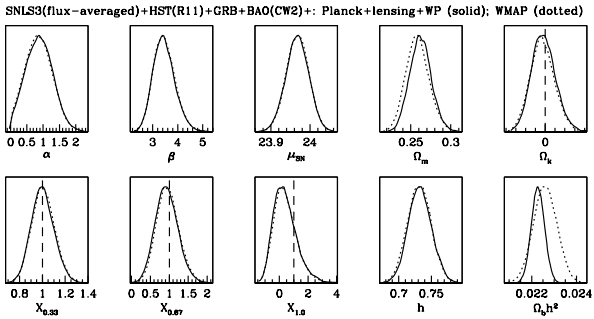,width=3.5in}\\
\vspace{-1.7in}
\caption{\label{fig:Xn3_pdf}\footnotesize%
The marginalized probability distributions for
$\{X_{0.33}, X_{0.67}, X_{1.0}, \Omega_m, \Omega_k, h, \omega_b,
\alpha, \beta, {\cal M}\}$, for SNe+$H_0$+GRB+GC(CW12), combined
with Planck+lensing+WP (solid) and WMAP9 (dotted) data respectively.
}
\end{figure}

\begin{figure} 
\psfig{file=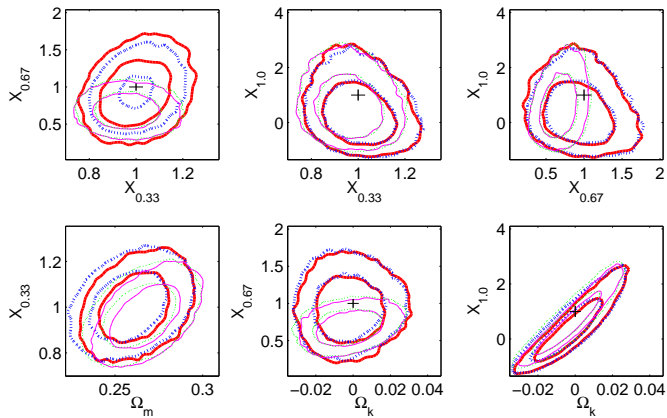,width=3.5in}\\
\caption{\label{fig:Xn3_2D}\footnotesize%
The joint 68\% and 95\% confidence contours for parameters of interest 
for SNe+$H_0$+GRB+GC(CW12), combined
with Planck+lensing+WP (solid) and WMAP9 (dotted) data respectively.
The thin solid and dotted contours also include GC data from BOSS.
}
\end{figure}

\begin{figure} 
\psfig{file=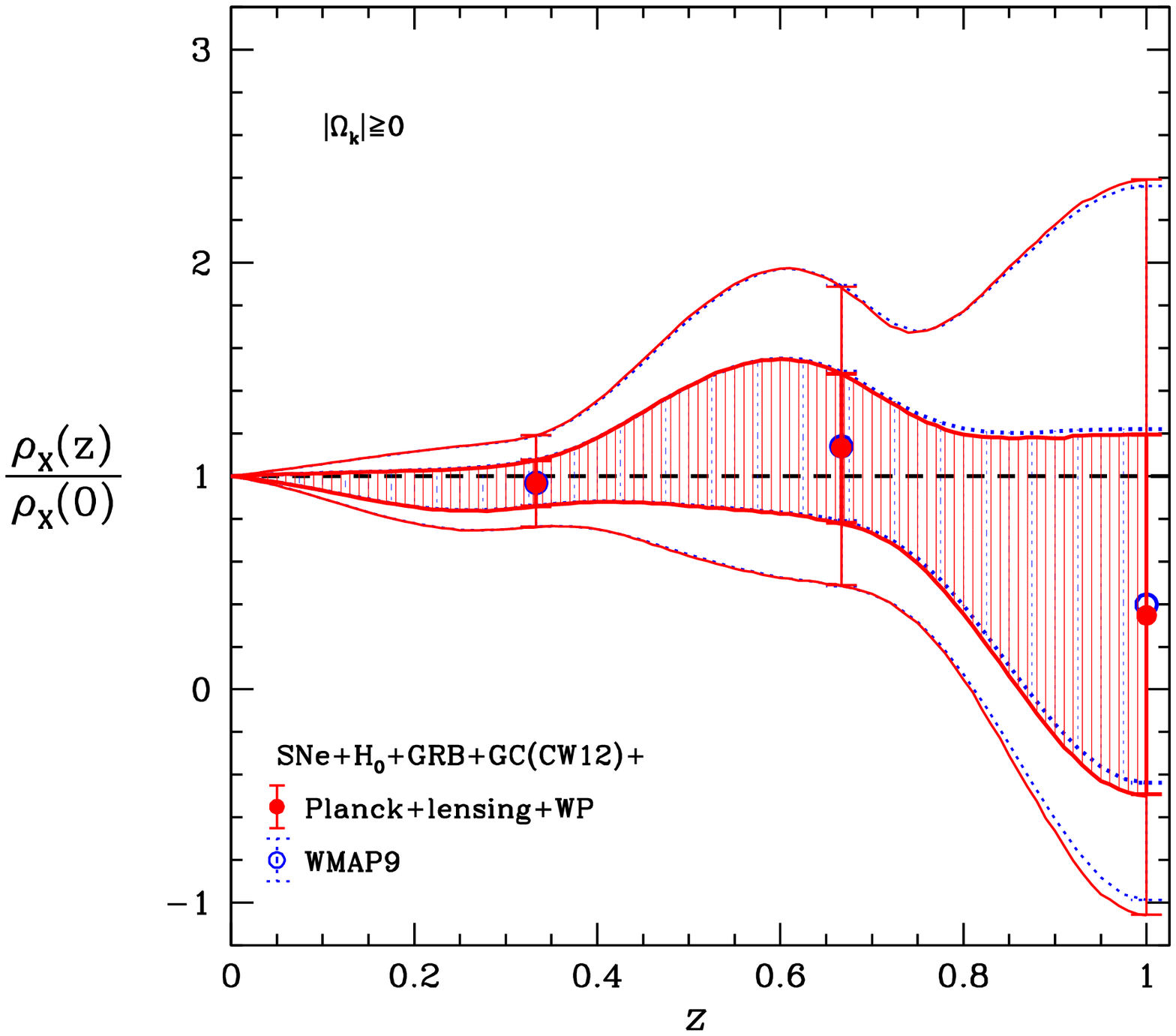,width=3.5in}\\
\caption{\label{fig:Xz_allbutBOSS}\footnotesize%
The 68\% and 95\% confidence constraints for $X(z)$ (cubic-splined from
$\{X_{0.33}, X_{0.67}, X_{1.0}\}$),
for SNe+$H_0$+GRB+GC(CW12), combined
with Planck+lensing+WP (solid) and WMAP9 (dotted) data respectively.
}
\end{figure}

\begin{figure} 
\psfig{file=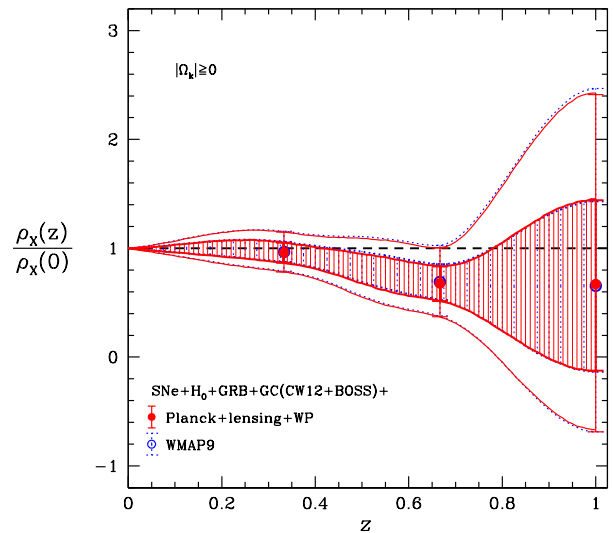,width=3.5in}\\
\caption{\label{fig:Xz_all}\footnotesize%
The 68\% and 95\% confidence constraints for $X(z)$ (cubic-splined from
$\{X_{0.33}, X_{0.67}, X_{1.0}\}$),
for SNe+$H_0$+GRB+GC(CW12+BOSS), combined
with Planck+lensing+WP (solid) and WMAP9 (dotted) data respectively.
}
\end{figure}

Figs.\ref{fig:Xn3_pdf}-\ref{fig:Xz_all} summarize our constraints
on $X(z)$ parametrized by its value at $z=1/3$, $2/3$, and 1.
Planck data give very similar results as WMAP9 data on $X(z)$, even although 
Planck data favor higher $\Omega_m$. However, note that
adding Planck priors leads to a marginal inconsistency with a cosmological
constant in a flat universe (see bottom right panel in Fig.\ref{fig:Xn3_2D}).

Adding BOSS data has a more significant impact: it shifts the
value of $X_{0.67}$ away from 1 at $\sim\,$1.5$\sigma$
(see Fig.\ref{fig:Xn3_2D} and Fig.\ref{fig:Xz_all}), independent
of cosmic curvature.

\subsection{Constraints on a linear dark energy equation of state}

We have studied the constraints on both ($w_0,w_a$) and ($w_0, w_{0.5}$),
as these have different base parameters (assumed to have flat priors).
In order to compare with previous work, and to display the impact of
replacing WMAP9 priors with Planck+lensing+WP priors, we do not include
GC data from BOSS in this comparison.

\begin{figure} 
\psfig{file=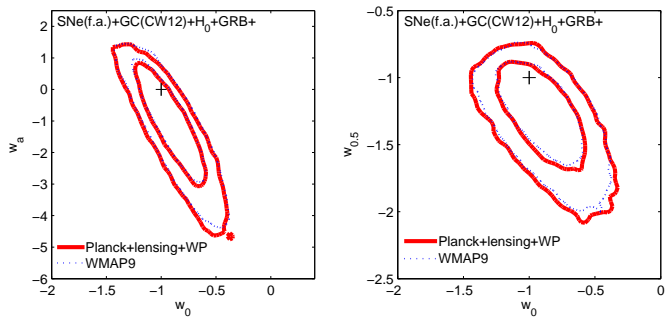,width=3.5in}\\
\caption{\label{fig:w0w1_allbutBOSS}\footnotesize%
The joint 68\% and 95\% confidence contours for ($w_0,w_a$) (left panel) and
($w_0, w_{0.5}$) (right panel), 
for SNe+$H_0$+GRB+GC(CW12), combined
with Planck+lensing+WP (solid) and WMAP9 (dotted) data respectively.
}
\end{figure}

\begin{figure} 
\psfig{file=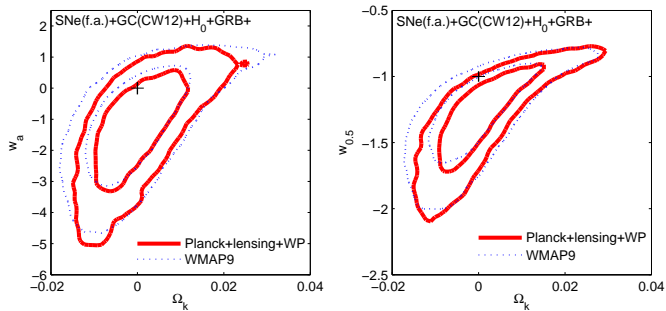,width=3.5in}\\
\caption{\label{fig:omegakw1_allbutBOSS}\footnotesize%
The joint 68\% and 95\% confidence contours for ($\Omega_k,w_a$) (left panel) and
($\Omega_k, w_{0.5}$) (right panel), 
for SNe+$H_0$+GRB+GC(CW12), combined
with Planck+lensing+WP (solid) and WMAP9 (dotted) data respectively.
}
\end{figure}

Fig.\ref{fig:w0w1_allbutBOSS} shows the joint 68\% and 95\% confidence contours for 
($w_0,w_a$) (left panel) and ($w_0, w_{0.5}$) (right panel), 
for SNe+$H_0$+GRB+GC(CW12), combined with Planck+lensing+WP (solid) and 
WMAP9 (dotted) data respectively. 
Fig.\ref{fig:omegakw1_allbutBOSS} shows the corresponding joint 68\% and 95\% 
confidence contours for ($\Omega_k,w_a$) (left panel) and
($\Omega_k, w_{0.5}$) (right panel), for the same data and with
the same line types. Fig.\ref{fig:w0w1_allbutBOSS} indicates that 
Planck priors do not have a significant impact
on the constraints on a linear dark energy equation state;
this means that the DETF FoM remains approximately the same compared
to that found by \cite{Wang12CM} using WMAP7 priors (using WMAP9 priors gives
similar results as using WMAP7 priors).
It is interesting to note that the right panel of Fig.\ref{fig:omegakw1_allbutBOSS}
shows that the combined data with Planck priors rule out $w_{0.5}=-1$
and a flat universe at $\sim\,1.5\sigma$. Planck data favor a small but
positive $\Omega_k$ (i.e., a slightly open universe).

\subsection{Constraints on a constant dark energy equation of state}

In order to understand better the difference between the Planck+lensing+WP
priors and the WMAP9 priors, we now study a constant dark energy equation
of state, for the minimal combination of SNe+$H_0$ data with the CMB priors.

\begin{figure} 
\psfig{file= 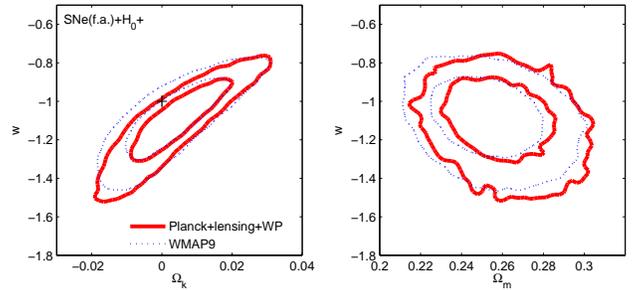,width=3.5in}\\
\caption{\label{fig:w_CMB_SNe_HST}\footnotesize%
The joint 68\% and 95\% confidence contours for ($\Omega_k, w$) (left panel) and
($\Omega_m,w$) (right panel), for SNe+$H_0$, combined
with Planck+lensing+WP (solid) and WMAP9 (dotted) data respectively.
}
\end{figure}

Fig.\ref{fig:w_CMB_SNe_HST} shows the joint 68\% and 95\% confidence contours for ($\Omega_k, w$) (left panel) and
($\Omega_m,w$) (right panel), for SNe+$H_0$, combined
with Planck+lensing+WP (solid) and WMAP9 (dotted) data respectively.
Here, we see even more clearly the trend of Planck+lensing+WP priors favoring a small but postive
$\Omega_k$ (a slightly open universe), and a somewhat higher $\Omega_m$,
compared to WMAP9 priors. Again, we find that adding Planck priors
to other data leads to a marginal inconsistency with a cosmological
constant in a flat universe.

\section{Discussion and Summary}

We have derived the distance priors from Planck first data release, in the
form of the mean values and covariance matrix of $\{R, l_a, \Omega_b h^2, n_s\}$,
which give an efficient summary of Planck data in the context of dark energy
constraints. As a test of the accuracy of this approach, we compare the
constraints on a constant dark energy equation of state $w$ in a flat universe
using Planck+lensing+WP data combined with SNLS SNe (no flux-averaging) in the form
of MCMC chains from the Planck data archiv, with the Planck+lensing+WP data
summarized by $\{R, l_a, \Omega_b h^2\}$ combined with the same SN data.
We find $w=-1.10 \, (-1.18, -1.02)$ for Planck+lensing+WP data summarized by
$\{R, l_a, \Omega_b h^2\}$, in excellent agreement with
$w=-1.12 \,(-1.18, -1.05)$ from the full Planck+lensing+WP data.
Not surprisingly, the full CMB data give slightly tighter constraints,
but the difference is not statistically significant.

We have used constraints on $\{R, l_a, \Omega_b h^2\}$ from Planck data
in combination with other data to probe dark energy in a conservative
geometric approach.
We have considered three different dark energy parametrizations:
(1) Dark energy density $X(z)=\rho_X(z)/\rho_X(0)$, parametrized by splining
its values at $z=1/3$, 2/3, and 1.0.
(2) Dark energy equation of state linear in the cosmic scale factor,
$w_X(z)=w_0+w_a(1-a)$, as well as the alternative parametrization with
much less correlated parameters $w_0$ and $w_{0.5}$, 
$w_X(z)=w_0(3a-2)+ 3w_{0.5}(1-a)$.
(3) Constant dark energy equation of state $w_X(z)=w$.

In addition to CMB priors, we used SNe compiled by Conley et al. (2011) \cite{Conley11},
flux-averaged to reduce systematic errors, the Hubble constant measurement by
Riess et al. (2011) \cite{Riess11}, the GRB data as summarized in Wang (2008) \cite{Wang08c},
and GC data from SDSS DR7 derived by Chuang \& Wang (2012) \cite{CW12}.
We have chosen to use the same data as in \cite{Wang12CM}, except for the
CMB priors (Planck+lensing+WP and WMAP9, versus WMAP7), in order to compare
the impact of the various CMB priors.
For completeness, we have added the GC data from BOSS derived by Chuang et al. (2013) \cite{C13}
in constraining $X(z)$.

We find that when dark energy density is allowed to be a free
function, current data (excluding GC data from BOSS, with either Planck+lensing+WP or WMAP priors) 
are fully consistent with a cosmological constant and a flat universe at 95\% C.L., but deviate from 
a cosmological constant in a flat universe at $\sim\,$68\% C.L. (see Fig.\ref{fig:Xn3_2D}).
The addition of BOSS data leads to a deviation from a cosmological constant at the redshift near 
that of the BOSS data at $>\,68\%$ confidence independent of cosmic curvature
(see Fig.6).
In general, adding Planck+lensing+WP priors leads to a preference for a 
small positive $\Omega_k$ (i.e., a slightly open universe) for a cosmological constant, 
or a flat universe with dark energy deviating from a cosmological constant,
compared to adding WMAP9 priors (see Fig.\ref{fig:Xn3_2D}).

When dark energy equation of state is assumed to be a linear function in 
the cosmic scale factor $a(t)$, the dark 
energy constraints depend on the base parameters used, the highly correlated $\{w_0,w_a\}$, or
the much less correlated $\{w_0,w_{0.5}\}$. The constraints on $\{w_0,w_a\}$ 
are consistent with a cosmological constant and a flat universe for
both Planck+lensing+WP and WMAP9 priors, while
that on $\{w_0,w_{0.5}\}$ are marginally inconsistent with a cosmological constant
in a flat universe (see Fig.\ref{fig:w0w1_allbutBOSS} and
Fig.\ref{fig:omegakw1_allbutBOSS}) for Planck+lensing+WP priors, 
similar to our findings in the $X(z)$ case.

We find that the above trend in the Planck+lensing+WP versus WMAP9
comparison becomes more pronounced when we assume 
a constant dark energy equation of state.
Here we only combine CMB priors with a minimal set of other data,
SNe and $H_0$, to highlight the difference between Planck+lensing+WP
and WMAP9 priors. We find that adding the Planck+lensing+WP priors
to the SNe and $H_0$ measurement leads to an inconsistency with 
a cosmological constant in a flat universe at $>\,68$\% confidence
level (see Fig.\ref{fig:w_CMB_SNe_HST}).

To conclude, we find that Planck distance priors are significantly tighter
than those from WMAP9 (see Figs.\ref{fig:l_a}-\ref{fig:obh2}).
However, adding Planck distance priors does not lead to significantly
improved dark energy constraints using current data, compared to
adding WMAP9 distance priors. This is because Planck data appear to
favor a higher matter density and lower Hubble constant \cite{planckXVI}, 
in tension with most of the other current cosmological data sets.

In order to understand the nature of dark energy, we will need
to improve our understanding of the systematic uncertainties of
all data used. Future dark energy measurements from 
space \cite{jedi,Cimatti09,euclid,wfirst} that
minimize systematic uncertainties by design will enable us
to make dramatic progress in our quest to shed light on dark energy.

\bigskip

{\bf Acknowledgments}
We acknowledge the use of Planck data archiv and CosmoMC.
This work is supported in part by DOE grant DE-FG02-04ER41305.
     
%\end{document}


\begin{thebibliography}{}

\bibitem[Riess et al.~(1998)]{Riess98}
Riess, A. G, {\etal}, 1998, Astron. J., 116, 1009

\bibitem[Perlmutter et al.~(1999)]{Perl99} 
Perlmutter, S. {\etal}, 1999, ApJ, 517, 565

%reviews:

\bibitem{Copeland06}
Copeland, E.~J., Sami, M., Tsujikawa, S., IJMPD, 15 (2006), 1753

\bibitem{Ruiz07}
Ruiz-Lapuente, P., Class. Quantum. Grav., 24 (2007), 91 

\bibitem{Ratra07}
Ratra, B., Vogeley, M.~S., arXiv:0706.1565 (2007)

\bibitem{Frieman08}
Frieman, J., Turner, M., Huterer, D., ARAA, 46, 385 (2008)

\bibitem{Caldwell09}
Caldwell, R. R., \& Kamionkowski, M., arXiv:0903.0866

\bibitem{Uzan09}
Uzan, J.-P., arXiv:0908.2243

\bibitem{Wang10}
Wang, Y., {\it Dark Energy}, Wiley-VCH (2010)

\bibitem{Li11}
Li, M., et al., 2011, arXiv1103.5870	

\bibitem{Weinberg12}
Weinberg, D. H.; et al., Physics Reports, in press, arXiv:1201.2434

%BAO:

\bibitem[Blake \& Glazebrook(2003)]{BG03}
Blake, C.; Glazebrook, K., 2003, ApJ, 594, 665

\bibitem[Seo \& Eisenstein(2003)]{SE03}
Seo, H; Eisenstein, D J 2003, ApJ, 598, 720

%WL:
	
\bibitem[Hu(2002)]{Hu02}	
Hu, W.,	2002, PRD, 66, 083515 
% (arXiv:astro-ph/0208093)
	
\bibitem[Jain \& Taylor(2003)]{Jain03}
Jain, B. \& Taylor, A. PRL, 91, 141302 (2003)

%CMB
\bibitem[Komatsu et al.(2011)]{Komatsu11}
Komatsu, E., et al. 2011, Astrophys.J.Suppl., 192, 18

%H_0:
\bibitem[Riess et al.~(2011)]{Riess11}
Riess, A. G, {\etal}, 2011, ApJ, 730, 119  
%arXiv:1103.2976 

%GRBs:

\bibitem{Amati02}
Amati, L., et al. 2002, A\&A, 390, 81 

\bibitem[Bloom, Frail, \& Kulkarni(2003)]{Bloom03}
Bloom, J. S., Frail, D. A., \& Kulkarni, S. R. 2003, ApJ, 594, 674

\bibitem{Schaefer03}
Schaefer, B. E., 2003, ApJ, 583, L71

% geometric/growth degeneracy:

\bibitem[Wang(2008a)]{Wang08a}
Wang, Y., Journal of Cosmology and Astroparticle Physics, 05, 021 (2008).
%JCAP05, 021
%arXiv:0710.3885 [astro-ph] 
%Differentiating dark energy and modified gravity with galaxy redshift surveys

\bibitem[Simpson \& Peacock(2010)]{Simpson10}
Simpson, F., \& Peacock, J.A. 2010, Phys Rev D, 81, 043512  

% CMB:

\bibitem[Page(2003)]{Page03}
Page, L., et al. 2003, ApJS, 148, 233 

\bibitem[Wang \& Mukherjee(2007)]{WangPia07}
Wang, Y., \& Mukherjee, P., PRD, 76, 103533 (2007)
%astro-ph/0703780

\bibitem{Li08}
Li, H., et al., ApJ, 683, L1 (2008)

\bibitem{Wang08b}
Wang, Y., 2008b, Phys. Rev. D 77, 123525 
%Figure of Merit for Dark Energy Constraints from Current Observational Data

\bibitem[Hu \& Sugiyama(1996)]{Hu96}
Hu, W., \& Sugiyama, N. 1996, ApJ, 471, 542

\bibitem[Eisenstein \& Hu(1998)]{EisenHu98}
Eisenstein, D. \& Hu, W. 1998, ApJ, 496, 605

\bibitem[Ade et al.(2013)]{planckXVI}
Ade, P. A. R., et al., arXiv:1303.5076
% Planck 2013 results. XVI. Cosmological parameters 

\bibitem[Bennett et al.(2012)]{Bennett12}
Bennett, C. L., et al.,  arXiv:1212.5225
% Nine-Year Wilkinson Microwave Anisotropy Probe (WMAP) Observations: Final Maps and Results 

\bibitem[Wang, Chuang, \& Mukherjee(2012)]{Wang12CM}
Wang, Y.; Chuang, C.-H.; \& Mukherjee, P., Phys. Rev. D 85, 023517 (2012)
% 	arXiv:1109.3172 
% A Comparative Study of Dark Energy Constraints from Current Observational Data

% GC:
\bibitem{CW12}
Chuang, C.-H.; and Wang, Y., MNRAS, 426, 226 (2012)
% 	arXiv:1102.2251
% Measurements of H(z) and D_A(z) from the Two-Dimensional Two-Point Correlation Function of Sloan Digital Sky Survey Luminous Red Galaxies

% SNe:

\bibitem{Conley11}
Conley, A., et al., 2011, Astrophys.J.Suppl., 192, 1

\bibitem[Wang(2000)]{Wang00}
Wang, Y., ApJ 536, 531 (2000)

\bibitem[Wang \& Mukherjee(2004)]{WangPia04}
Wang, Y., \& Mukherjee, P. 2004, ApJ, 606, 654

\bibitem[Wang(2005)]{Wang05}
Wang, Y., JCAP, 03, 005 (2005)
%astro-ph/0406635

\bibitem{Hui06}
Hui, L., \& Greene, P.B., PRD, 73, 123526 (2006) 
%arXiv:astro-ph/0512159
%"Correlated Fluctuations in Luminosity Distance and the (Surprising) 
%Importance of Peculiar Motion in Supernova Surveys"

\bibitem{Clarkson11}
Clarkson, C., et al., arXiv:1109.2484v2 
% "(Mis-)Interpreting supernovae observations in a lumpy universe"

% GC:

\bibitem[Chuang et al.(2013)]{C13}
Chuang, C.H., et al., arXiv:1303.4486
% The clustering of galaxies in the SDSS-III Baryon Oscillation Spectroscopic Survey: single-probe measurements and the strong power of normalized growth rate on constraining dark energy 



% GRB:

\bibitem{Wang08c}
Wang, Y., 2008b, PRD, 78, 123532 
%Model-Independent Distance Measurements from Gamma-Ray Bursts and Constraints 
%on Dark Energy

\bibitem{Schaefer07}
Schaefer, B. E., 2007, ApJ, 660, 16

\bibitem{Butler07}
Butler, N.R. et al., ApJ, 2007, 671, 656

\bibitem{Butler09}
Butler, N.R. et al., ApJ, 2009, 694, 76  

\bibitem{Petrosian09}
Petrosian, V.; Bouvier, A.; Ryde, F., arXiv:0909.5051

\bibitem{Shahmoradi11}
Shahmoradi, A., \& Nemiroff, R.J., MNRAS, 2011, 411, 1843                   
% "The possible impact of gamma-ray burst detector thresholds on                    
% cosmological standard candles"          

% other:

\bibitem[Wang \& Garnavich(2001)]{WangGarnavich}
Wang, Y., and Garnavich, P. 2001, ApJ, 552, 445

\bibitem[Wang \& Tegmark(2004)]{WangTegmark04}
Wang, Y., \& Tegmark, M. 2004, Phys. Rev. Lett., 92, 241302 

\bibitem[Wang \& Freese(2006)]{WangFreese06}
Wang, Y., \& Freese, K. 2006, Phys.Lett. B632, 449
(astro-ph/0402208)

\bibitem[Linde(1987)]{Linde87}
Linde, A. D., ``Inflation And Quantum Cosmology,'' in
{\it Three hundred years of gravitation}, (Eds.: Hawking, S.W. and Israel, W.,
Cambridge Univ. Press, 1987), 604-630.

\bibitem{WangLinde04}
Wang, Y.; Kratochvil, J. M.; Linde, A.; \& Shmakova, M., 
JCAP 0412 (2004) 006
% arXiv:astro-ph/0409264v2

% parametrizations
\bibitem[Chev01(2001)]{Chev01}
Chevallier, M., \& Polarski, D. 2001, Int. J. Mod. Phys. D10,
213


\bibitem[Lewis02(2002)]{Lewis02}
Lewis, A., \& Bridle, S. 2002, PRD, 66, 103511


% future:

\bibitem{jedi}
Crotts, A. et al., 2005, astro-ph/0507043
% Joint Efficient Dark-energy Investigation (JEDI): a Candidate Implementation of the NASA-DOE Joint Dark Energy Mission (JDEM)

\bibitem{Cimatti09}
Cimatti, A., et al., Experimental Astronomy, 23, 39 (2009)
% SPACE: the spectroscopic all-sky cosmic explorer

\bibitem{euclid}
Laureijs, R.; et al., 2011, arXiv1110.3193
%  Euclid Definition Study Report

\bibitem{wfirst}
Green, J.; et al., 2012, arXiv1208.4012
% Wide-Field InfraRed Survey Telescope (WFIRST) Final Report



\end{thebibliography}
\end{document}